# LπCT: A logic security analysis framework for protocols

Fusheng Wu [a], Jinhui Liu [b], Yanbing Li[c] and Mingtao Ni [d*]

[a] Guizhou Key Laboratory of Economic System Simulation, Guizhou University of Finance and Economics, Guiyang, China
[b] School of Computer science, Northwestern Polytechnical University, Xian, China
[c] College of Artificial Intelligence, Nanjing Agricultural University, Nanjing, Cina
[d] School of Computer, Leshan Normal University, Leshan, China

**Abstract :** The π-calculus is a basic theory of mobile communication based on the notion of interaction, which, aimed at analyzing and modelling the behaviors of communication process in communicating and mobile systems, is widely applied to the security analysis of cryptographic protocol's design and implementation. But the π-calculus does not provide perfect logic security analysis, so the logic flaws in the design and the implementation of a cryptographic protocol can`t be discovered in time. The aim is to analyze whether there are logic flaws in the design and the implementation of a cryptographic protocol, so as to ensure the security of the cryptographic protocol when it is encoded into a software and implemented. This paper introduces logic rules and proofs, binary tree and the KMP algorithm, and proposes a new extension the π-calculus theory, a logic security analysis framework and an algorithm. This paper presents the logic security proof and analysis of TLS1.3 protocol's interactional implementation process. Empirical results show that the new extension theory, the logic security analysis framework and the algorithm can effectively analyze whether there are logic flaws in the design and the implementation of a cryptographic protocol. The security of cryptographic protocols depends not only on cryptographic primitives, but also on the coding of cryptographic protocols and the environment in which they are implemented. The security analysis framework of cryptographic protocol implementation proposed in this paper can ensure the security of protocol implementation.

## 1.Introduction

In the real world, information can be collected, transmitted, stored and processed through interactional behavior among entities, and this way has been migrated to the virtual world. So far, with the environment of running a software collaboratively becoming more and more complex, the software implementation has been involved in interactional behavior. For example, entities can exchange information in the cyberspace; the π-calculus is based on the original concepts of "interaction", as distinct from Turing machines and registers, which are based on read/write access media, as well as recursive equations and λ-calculus, which are based on mathematical functions. In the mobile communication system, the π-calculus is the basic theory processing in the virtual space, which is used to explore the relationship between process position and process link in virtual process space, that is, the position of a moving process is determined by its links with other processes.

Therefore, the π-calculus is applied to the study of process states transition in mobile communication system, which not only formalizes the behavior of process states transition, but also formalizes data structures, classes, objects and functions in code design. It can be seen that the π-calculus is applicable not only to the static study of program design in mobile communication systems, but also to the dynamic detection of the process when the program (source code) is implemented. Therefore, it is widely used in the security analysis of cryptographic protocol design and implementation, model checking and distributed system security and other fields.

The security of software implementation is the premise of information security ensuring the interactional transmission, processing and storage, and it is a long-term challenge faced by academia and industry. It is one of the basic methods to study whether there are logic flaws in software design and implementation to reduce the occurrence of vulnerabilities and avoid attacks by exploiting vulnerabilities.

The π-calculus theory is applicable to the security analysis of software design and implementation, such as key exchange protocol, key agreement protocol and TLS series of protocols. From the perspective of application, cryptographic protocol design and implementation belongs to the scope of software design and implementation, so it can't escape from logic flaws, which are insidious and difficult to find. When designing and implementing cryptographic protocols, these flaws often lead to the generation of vulnerabilities. This situation not only appears in the design and implementation of cryptographic protocols, but also in other aspects of software design and implementation. the π-calculus is widely used in the study of states transition in the design and implementation of cryptographic protocols, carrying out formal analysis, such as interaction modeling between concurrent processes, structural congruence of processes and strong bisimulation and observation equivalence. If the π-calculus can be used to analyze whether there are logic flaws in the design and implementation of cryptographic protocols, it can reduce the frequency of vulnerabilities and attacks using vulnerabilities during the implementation of cryptographic protocols. In other words, with the π-calculus, the logic flaws from the cryptographic protocol design and implementation will be found, which, by promoting the security of cryptographic protocol design and implementation, will reduce insecure events.

How to use the π-calculus to find logic flaws in the design and implementation of cryptographic protocols? This is a major problem in software design and implementation.

According to the basic theory of the π-calculus, logic rules and proofs are the basic criteria of logic security analysis, which can be used for logic analysis and logic reasoning of natural language semantics or entity transition behavior. The natural language semantics and entity transition in the real world are abstracted into symbols, and then logic rules and proofs are used for modeling, logic reasoning and formal analysis of the natural language semantics and entity transition(Michael HuthMark,2004), so as to discover whether there are logic flaws in the entity transition of software. In conclusion, the π-calculus can be used to find logic flaws. If the π-calculus is extended, by adding new logic rules and proofs, it can be used to find and analyze the logic flaws in the software design and implementation .

To analyze whether there are logic flaws in the design and implementation of cryptographic protocols, this paper, by adding logic rules and logic proofs, the extends π-calculus theory. The extended theory can be applied to the design (idea and concrete steps) and implementation (coding and environmental testing) of cryptographic protocols, to reduce the vulnerabilities from cryptographic protocol implementation and avoid the attacks from vulnerabilities, so as to ensure the security and reliability of software implementation.

The main contributions of this paper:

1. Propose extended the π-calculus theory by adding new logic rules and proofs.

2. Propose node states transition binary tree and corresponding event binary tree on the base of extended the π-calculus theory.

3. Propose a logic security analysis framework for interactional implementation of cryptographic protocols.

4. Prove whether there are logic flaws in the cryptographic protocol implementation.

**2.Related works**

The classical logic (Monin J F,et al.,2003) mainly conducts logic analysis of natural language based on logic rules and logic proofs. It is widely used in the fields such as text semantic analysis of natural language, integrated circuit design, artificial intelligence, cyberspace security, and so on. In the field of computer science, mathematical logic (MICHAEL HUTH,et al.,2004) is widely used in the fields, such as model check, program verification, symbolic analysis, providing theory support for the security and stability of software design and implementation.

"Interactional" behavior is a universal phenomenon in the real world, and also widely exists in the virtual world, such as information transmission, data processing and human-computer interaction. Based on process interactional behavior, the π-calculus theory has been established and widely used in the fields of mobile communication system and cyberspace security, such as the security analysis of cryptographic protocol behavior, model check, axiomatic semantic security verification. Literature (David Baelde,et al.,2021) proposes a framework and an Interactional proof program, which allows the mechanized proof for arbitrary number of sessions of security protocols in the calculation model, and develops a meta-logic and develops a meta-logic and a proof system to derive security attributes.

Similarly, in order to solve the problem of multi-core programming, literature (Petersen L,et al.,2010), on the base of π-calculation theory, proposes to describe the security attribute of interaction behavior with only two semantic rules in multi-core programming, and reinterprets part of the detecting terminal problem. In the study of the symbolic representation of double similarity, in order to overcome the imperfection of expression of the π-calculus, literature (Ross Horne,2018) proposes the complete symbolic mutual simulation method of the π-calculus, and the simulation concept of symbolic double similarity is perfected, so that the double similarity work of the π-calculus theory is extended and studied (Cheval V,et al.,2018)( Bodei C,et,al.,2020). In the study of cryptographic protocol security analysis based on the π-calculus theory, literature (Bruno Blanchet,2016) discusses how to apply the π-calculus theory in model check and cryptographic protocol implementation verification, which makes it possible to apply the π-calculus theory in ProVerif verification tool.

In the research of the π-calculus semantics, literature (Thomas Troels,et al.,2019) proposes a stable the π-calculus with non-interleaved operation semantics, and deduces the dynamic topology of the π-calculus, or the causal relationship between objects. For more details of the semantic research of the π-calculus theory, please refer to literature (Kokke W,et al.,2018)( Niccolò Veltri,2020). In the mobile communication system, the communication between the static domain name and the channel message is carried out in an Interactional way. In the process of protocol communication, it is very important to ensure secure Interactional communication. Literature (Martín Abadi,et al.,2017) discusses the research of the π-calculus theory on the communication security of the mobile system. the π-calculus theory is applied not only to the behavioral security analysis in communication system processes, but also to the cyberspace physical system (CPS). Literature (Livinus Obiora Nweke,et al.,2021) extends the study of applying the π-calculus theory, and proposes a method , which can capture the behavior of CPS and simulate the behavior of adversaries. Besides, this method has embedded the attack defense tree (ADT) into the application of the π-calculus, so as to expand its defense capability by adding ADT sorting.

In the field of cyberspace security, the logic security analysis of software design and implementation is applied in order to find the logic flaws of software implementation and reduce the vulnerabilities and attacks exploiting vulnerabilities during software implementation. Aimed at the logic security analysis of development guide document in the system, literature (Yi Chen,et al.,2019) proposes to find the logic flaws in the joint payment system service through automatic document analysis method. This method effectively detects the potential logic flaws of development guide documents in the system, and thus finds five logic flaws in the joint payment system. These logic flaws leads to the payment vulnerabilities in the implementation of the joint payment system. In a joint payment system, an attacker can bypass payments or underpay.

With the help of machine learning and automatic reasoning, finding logic flaws in software (program source code) is the future direction of automatic document analysis. The bottom layer of software implementation is the program source code. If there are logic flaws in the design of source code (definition of variables, assignment and definition and call of functions, etc.), the security of software implementation will be directly affected. Literature (Edward J,et al,2018) proposes to use logic programming to recover C++ classes from compiled executable files. Literature (Edward J, et al.,2018) applies manual recognition of binary code to make up for the deficiencies of C++ program, and proposes a new method combining lightweight symbol analysis with Prolog logic reasoning system, so as to improve the recognition rate of C++ malicious code. Logic, language and security are the basis of software implementation security, and there is a strong causal relationship among them. For more details of relevant studies on logic and language, please refer to literature (Nigam, V,et al.,2020).

In the field of interaction behavior security analysis of cryptographic protocol, Hoare logic has been widely applied, such as model check and attribute security verification, etc. Hoare logic checks and verifies the formula syntax and judges the satisfiability of the formula. The satisfiability judgment of Horn logic clause has been widely applied in the field of cyberspace security. For example, source code level security analysis of Needham-Schroeder protocol in C language is first presented in literature (Jean Goubault-Larrecq,et al.,2005). This method, with Horne clause logic as a tool, introduces assertion method and assumes adversaries with the capability to have Dolev-Yao model, so as to analyze the security of the cryptographic protocol implementation at the source code level. But this method can not give the mutual authentication function of protocol implementation.

Proposed in literature (Jia Chen,et al.,2017), Quantitative Cartesian Hall logic (QCHL) is a program logic verification tool that is used in Java application programs to find side channel vulnerabilities arising when searching for resource utilization and find unknown side channel vulnerabilities. However, this tool can only implement two types of cost models, namely time and response, and cannot meet the applicability of cost models provided by other types of resources. In the research on axiomatic semantic verification of first-order logic, in order to improve the reasoning efficiency of module reading, literature (Anindya Banerjee, et al.,2018) proposes the reasoning principle of end-to-end module, and connects the implementation of modular method with clients. Aimed at the language research on dynamic variable allocation objects, Hoare logic is introduced in literature (Banerjee A,et al.,2019) to encapsulate variables and constants, and realize the coupling of the correlation between them. In the formal verification of smart contracts by blockchain, literature (Ribeiro M,et al.,2020), uses Hoare logic to describe the attributes of the program, proposes a proof system of program language, and discusses the robustness and (relative) integrity of the program. In addition to its extensive applications in classical computers, Hoare logic has also been extended to the field of quantum, such as the work in literature (Tao Liu,et al.,2016)( Li Zhou,et al.,2019).

Logic rules and proofs ensure the logic security of entity communication behavior and reduce or avoid the logic flaws in software design and implementation. In the π-calculus theory, the introduction of logic rules and proofs is a prerequisite to ensure the security of process interaction communication.

## 3.Extended the π-calculus theory

The π-calculus theory discuses the dynamic links among entity processes in the mobile communication systems, such as mobile phones, mobile communication location, collaborative operation of software, etc. Processes are linked logic. Logic rules are the basic theory to analyze mobile process behavior, and the key technology to analyze whether there are logic flaws in the software design and implementation. Due to limited length, this paper can only present some basic theoretical knowledge of the π-calculus. For further information, please refer to literature (Milner R,99)( Hennessy, M,2007).

3.1The basic theoretical knowledge of the π-calculus

1.The basic syntax of the π-calculus theory is represented by Backus-Naur Form (BNF).The action prefix of the π-calculus refers to sending a message (i.e., a name), receiving a message or executing a internal transfer. Its grammar is written as follows:

$\pi ::= x(y)$ Denotes x receives y

$\bar{x} < y >$ Denotes x sends y

$\tau$ Denotes invisible action ( internal response)

2.Set $P^\pi$ of the π-calculus process expression:

$P^\pi ::= \sum_{i \in I} \pi_i.P_i \mid P_1 \mid P_2 \mid new\ aP \mid !P$. $I$ is subscript of finitely. $\sum_{i \in I} \pi_i.P_i$ is called process sum.

3.Structural congruence：if two the π-calculus processes $P$ and $Q$ are structural congruence, they can transform with each other through the following rules :

①Change of bound names (alpha-conversion)

②Reordering of terms in a summation

③ $P \mid 0 \equiv P$、$P \mid Q \equiv Q \mid P$、$P \mid (Q \mid R) \equiv (P \mid Q) \mid R$

④ $new\ x(P \mid Q) \equiv P \mid new\ Q$ if $x \notin fn(P)$
$new\ x0 \equiv 0$、$new\ xy\ P \equiv new\ yx\ P$

⑤ $!\ P \equiv P \mid !P$

4.Process standard expression: $new\ \bar{a}(M_1 \mid \cdots \mid M_m \mid !Q_1 \mid \cdots \mid !Q_n)$. If every $M_i$ is non-empty sum form, every $Q_j$ is standard.

5.The labelled transition system: a labelled transition system of the action set $Act$ is a two-tuples $(Q,T)$ Here, $Q$ is a states set ; $T$ is a triple, which is called transition relation, $T \subseteq (Q \times Act \times Q)$.

6. Basic rules (also called response rule)
① $TAU : \tau.P + M \to P$ ;
② $REACT : (a.P + M) \mid (\bar{a}.Q + N) \to P \mid Q$ ;
③
$REACT' : (x(\vec{y}).P + M) \mid (\bar{x} < \vec{z} >.Q + N) \to \{\vec{z}/\vec{y}\}P \mid Q$ ;

④ $PAR : \dfrac{P \to P'}{P|Q \to P'|Q}$ ;

⑤ $RES : \dfrac{P \to P'}{newaP|Q \to newaP'}$ ;

⑥ $STRUCT : \dfrac{P \to P'}{Q \to Q'}$ if $P \equiv Q$, $P' \equiv Q'$.

7. Arden law: any given string sets $S$ and $T$, equation $X = S.X + T$ has a solution $X = S * X$, and $\varepsilon \notin S$ is the only solution of this equation.

3.2 The basic knowledge of logic theory

Logic rules are the tools of logic security analysis, while logic proofs are the process of logic security analysis. Logic definition and logic rules are the basis of logic security analysis. The basis theory is as follows:

**Definition 1 a calculus of deducing**: assume that the formula $\phi_1, \phi_2, \cdots \phi_n$ is the precondition, and that the formula $\psi$ the conclusion. By applying proof rules to the precondition $\phi_1, \phi_2, \cdots \phi_n$, more formulas will be derived. More proof rules are applied to these derived formulas, and then the final conclusion $\psi$ will be reached. In this way, the sequence is valid, which is denoted as:

$$\phi_1, \phi_2, \cdots \phi_n \mathrel{|\!-} \psi \qquad (3\text{-}1)$$

**Definition 2 the semantic entailment relation**: If, for all valuations in which all $\phi_1, \phi_2, \cdots \phi_n$ evaluate to T, $\psi$ evaluate to T as well, we say that

$$\phi_1, \phi_2, \cdots \phi_n \models \psi \qquad (3\text{-}2)$$

holds and call $\models$ t the semantic entailment relation.

**Definition 3 provable equivalence** : Let $\phi$ and $\psi$ be formulas of propositional logic. We say that $\phi$ and $\psi$ are provably equivalent iff the sequent $\phi \mathrel{|\!-} \psi$ and $\psi \mathrel{|\!-} \phi$ are valid; that is, there is a proof of $\psi$ from $\phi$ and another one going the other way around. As seen earlier, we denote that $\phi$ and $\psi$ are provably equivalent by $\phi \dashv\vdash \psi$.

**Definition 4 soundness and Completeness**: Let $\phi_1, \phi_2, \cdots \phi_n, \psi$ be formulas of propositional logic. Then $\phi_1, \phi_2, \cdots \phi_n \models \psi$ is holds iff the sequent, $\phi_1, \phi_2, \cdots \phi_n \models \psi$ is valid.

3.3 Extended the π-calculus theory

Based on the interactional concept, the π-calculus is used to discuss the behavior trace of the process migration in the virtual space. According to classical automata theory, the π-calculus theory is essentially a principle of states transition, but in the communication systems, the π-calculus theory does not consider the logic security relation between the premises and the conclusions of states transition, nor does it consider the logic relationship among the correspondence events of states transition (the protocol has performed some other events before performing an event). As a result, the π-calculus lacks logic reasoning and logic provable security when analyzing the process movement.

In order to ensure the logic and security of the process movement in the virtual space, this paper, based on the π-calculus theory, introduces the premises of process states transition and logic reasoning and proof, and extends the research on the π-calculus theory. The extended the π-calculus theory has been proposed, which is integrated with logic rules and proofs ( $L\pi CT$ for short). $L\pi CT$ is applied to the security analysis of the design and implementation of cryptographic protocol and to the mobile communication systems, to analyze the abnormal detection from the interactional communication behavior of cryptographic protocols. Therefore, the security of the software implementation is ensured.

1. $L\pi CT$

The idea of $L\pi CT$ : logic rules are integrated with logic proof and the π-calculus theory in communication and mobile systems, to discuss the logic security relationship between the preconditions and the conclusions of states transition of cryptographic protocol. According to the π-calculus theory, cryptographic protocols implementation must consider the preconditions for the states transition, the transition process and the logic security of the states transition conclusions, so as to analyze whether there are logic flaws in cryptographic protocols implementation. The basic definitions of $L\pi CT$ are given below.

**Definition 5 the labelled transition system of $L\pi CT$** : according to the mobility and automaton theory of process link in the virtual space, let the set of actions of process states transition is $Act$. A labelled transition system of $Act$ is a ternary relation $(L,Q,T)$. Here, $L$ is indicated as logic rules of states transitions. $Q$ is indicated as a states set. $T$ is indicated as a labelled transition relation, denoted as : $T \subseteq (Q \times Act \times Q)$.

**Definition 6 the preconditions of $L\pi CT$** : according to $L\pi CT$, $q$ and $a$ satisfy the validity and soundness of the sequent of a node. That is, $q$ and $a$ satisfy logic $L$. $q$ and $a$ are deduced by logic rules and proof. Here, $q$, called the premises of states transition, can be deduced by logic $L$: $q_1,q_2\cdots q_n \mathrel{|\!-} q, n \in \mathbb{N}$ and $a_1,a_2\cdots a_n \mathrel{|\!-} a, n \in \mathbb{N}$, which are also valid sequent, and satisfy reasonability $L$.

**Definition 7 the conclusion of $L\pi CT$** : according to $L\pi CT$, the sequent, of $q$ and $a$ the logic entailment and semantic entailment of the sequent of $q$ and $a$. $q'$ is called the conclusion of states transition.

**Definition 8 the corresponding events of $L\pi CT$** : before the event $E$ occurs, some other events $E'$ must have occurred . That is , if $E'$ are preconditions for $E$ to occur, $E'$ are corresponding events of $E$ . These phenomena generally exists in the real world. For example, before a car runs, the preconditions must be satisfied: a tank with oil, normal operation of components, and so on. Before a mobile phone works, the preconditions must be satisfied : charged power, free from arrearage, and so on.

2. The basic logic rules of $L\pi CT$ ( or $L\pi CT$ response rules)

① $LTAU : l.\tau.P + M \to P$ ;

② $LREACT : (l.a.P + M) | (l.\bar{a}.Q + N) \to P | Q$ ;

③ $LREACT': \dfrac{(l.x(\vec{y}).P+M) \,|\, (l.\bar{x}<\vec{z}>.Q+N) \to \{\vec{z}/\vec{y}\}P\,|\,Q}{}$ ;

④ $LPAR: \dfrac{l.P \to P'}{l.P\,|\,l.Q \to P'\,|\,Q}$ ;

⑤ $LRES: \dfrac{l.P \to P'}{l.newaP\,|\,l.Q \to newaP'}$ ;

⑥ $LSTRUCT: \dfrac{l.P \to P'}{l.Q \to Q'}$, if $l.P \equiv l.Q$ and $l.P' \equiv l.Q'$ ;

⑦ $LSOOM: \dfrac{e_1 \wedge e_2 \wedge \cdots \wedge e_n}{E}$

or $\dfrac{e_1 \vee e_2 \vee \cdots \vee e_n}{E}$ or $\dfrac{(\wedge\,|\,\vee)\prod_i^n e_i}{E}$ , response rules.

Here, $l \in L$ , and $L$ satisfies the logic entailment or semantic entailment of states transition. That is, $L$ satisfies (3-1) and (3-2). $\wedge, \vee$ and $|$ are imported into corresponding rules. $\dfrac{(\wedge\,|\,\vee)\prod_i^n e_i}{E}$ means that processes have executed some other events before executing event $E$. $(\wedge\,|\,\vee)\prod_i^n e_i$ means the continuous conjunction calculation of $e_i$, or disjunction calculation, or the mixed calculation of conjunction and disjunction.

By integrating logic rules and logic proofs into the π-calculus theory, the π-calculus theory is extended, which makes up for the logical analysis of the Interactional states transition of cryptographic protocols in communication and mobile systems, as well as the logic detection for the abnormal behaviors from cryptographic protocol implementation. As a result, the logic flaws in the design and implementation of cryptographic protocols will be found, so as to ensure the security of software implementation.

**4. $L\pi CT$ Logical Security Analysis Framework**

On the base of definition 8 and binary trees theory, Depth-first Logic Security Analysis Framework for interactional implementation of cryptographic protocols is proposed, called DFLSAF for short.

4.1 $L\pi CT$ Binary trees theory

On the base of $L\pi CT$ and responding event definitions, the binary tree $T_{s_i}$ is built for states transition nodes $\{S_i\,|\,i \in N\}$, as shown in figure 4-1.

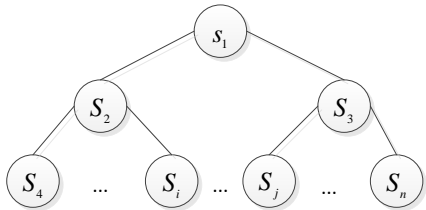

Figure 4-1 the binary tree of states transition nodes $S_i$

On the base of corresponding rules, the binary tree $T_{e_i}$ is built for the corresponding events of states transition nodes $\{S_i\,|\,i \in N\}$, which are ( $e_1, e_2, \cdots e_i \cdots e_j \cdots e_n$ , $2 \leq i,j,n$ , $i,j,n \in N$ ), as shown in figure 4-2.

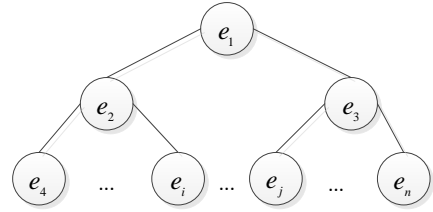

Figure 4-2 the binary tree of corresponding events $S_i$

According to the structure of the binary tree $T_{e_i}$, it is easy to search the nodes of the binary tree. Through logic calculation, judge whether the logic calculation for the corresponding events ( $e_1, e_2, \cdots e_i \cdots e_j \cdots e_n$ , $2 \leq i,j,n$ , $i,j,n \in N$ ) of states transition $S_i$ is true or not. If the logic calculation is true, the nodes $S_i$ of states transition are true and vice versa. If $S_i$ is true, there will not be logic flaws in the design and complementation of nodes $S_i$ in a cryptographic protocol and vice versa.

4.2 The description of DFLSAF algorithm

DFLSAF algorithm is described as follows:

1. In the protocol interactional procedure, the node set of states transition $\{S_i\,|\,i \in N\}$ constitutes the binary tree $T_{S_i}$.

2. Corresponding events ( $e_1, e_2, \cdots e_i \cdots e_j \cdots e_n$ , $2 \leq i,j,n$ , $i,j,n \in N$ ) of every node $S_i$ constitutes the binary tree $T_{ei}$.

3. Input states transition nodes $S_i$, and judge whether there is true logic calculation for the corresponding events $e_1, e_2, \cdots e_i \cdots e_j \cdots e_n$ , $2 \leq i,j,n$ , $i,j,n \in N$ of nodes $S_i$ through Breadth-First-Search algorithm ( BFS for short). If there is true logic calculation, DFLSAF will recursive call next node $S_{i+1}$. Assign $S_i = S_{i+1}$ , and execute 2 And 3. Otherwise, decide that there are logic flaws in the protocol Interactional implementation, and then quit the DFLSAF.

4. When all the states transition nodes $S_i$ are searched and judged, states transition $\{S_i\,|\,i \in N\}$ will be judged :

I. Whether states transition sequent $\{S_i\,|\,i \in N\}$ in the protocol interaction procedure satisfies partial ordering relation.

II. Whether a conclusion can be got through logic entailment calculation of logic rules $\dfrac{\phi\ \ \phi \to \psi}{\psi} \to i$ and $\dfrac{\phi \to \psi,\ \neg\psi}{\neg\phi}\ MT$ .

III. If I and II succeed, there are no logic flaws in the protocol Interactional implementation, and vice versa.

4.3 The DFLSAF Algorithm Pseudo-code Description

In the research of the logic security analysis of the design and implementation of cryptographic protocol, binary trees are introduced to express the logic calculation of corresponding events. With binary trees, DFLSAF algorithm can be converted into pseudocode, so it is easy to convert it into software.

| DFLSAF algorithm |
|---|
| step1: Input sets $\{S_i \mid i \in N\}$ and $\{e_i \mid i \in N\}$ ;
step2: Create the Binary Tree: $T_{s_i}$ ;
step3: Call BFS1( $S_i$ ) ;
{ Create the Binary Tree: $T_{e_i}$ ;
    if ( $S_i$ are not empty)
      { Call BFS2( $e_i$ );
        if ( $e_i$ is not empty)
          calculated: $(\wedge \mid \vee) \prod_{i}^{n} e_i$ ;
        if $(\wedge \mid \vee) \prod_{i}^{n} e_i$ is true
          BFS2( $e_{i+1}$ );
        else
          exit ;
      }
      BFS1( $S_{i+1}$ );
    else;
      exit ;
}
step4: if ( $\{S_i \mid i \in N\}$ are the partial order relation which is true);
    return 1;
  else
    return 0;
step5: if (step4==1&& Satisfied condition III )
    printf (" There are not logic flaws in the design
    and implementation of the cryptographic protocol
    or the cryptographic protocol doesn't suffer from attacks").
    else
     printf ("There are logic flaws in the design and
     implementation of the cryptographic protocol or the cryptographic
     protocol suffers from attacks").
     return 0; |

DFLSAF algorithm can be used to analyze whether there are any abnormal behaviors in the corresponding event sequent of the design or the implementation of a cryptographic protocol. In this way, logic flaws are quickly identified in the design of a cryptographic protocol, and abnormalities are located during the course of the implementation, so that the cause of the faults will be analyzed and the problems will be solved in time.

4.4 Creating the Environment for Implementing $L\pi CT$

$L\pi CT$ contains two different implementing environments: the ideal environment and the non-ideal environment. The steps of creating the environments for implementing $L\pi CT$ are as follows.

Definition 9 the ideal environment: there are not such communication environments assumed by Dolev-Yao model, in which there are no attacks(passive attacks or active attacks ). When sent and received, the information will be protected and read through encryption and decryption.

Definition 10 the non-ideal environment: there are communication environments assumed by Dolev-Yao model, in which there are attacks(passive attacks or active attacks ). Attackers have ability to get ciphertext and partial plaintext, or to attack the network.

Supposing: through DFLSAF, a cryptographic protocol gets sequent $\beta_1, \beta_2, \cdots \beta_i \cdots \beta_j \cdots \beta_n$ in the ideal environment and $\beta'_1, \beta'_2, \cdots \beta'_i \cdots \beta'_j \cdots \beta'_n$ in the non-ideal environment respectively.

Theorem 1: KMP algorithm are employed. If $\beta_1, \beta_2, \cdots \beta_i \cdots \beta_j \cdots \beta_n$ and $\beta'_1, \beta'_2, \cdots \beta'_i \cdots \beta'_j \cdots \beta'_n$ are matched successfully, there are not logic flaws in $\beta'_1, \beta'_2, \cdots \beta'_i \cdots \beta'_j \cdots \beta'_n$ . Therefore, the cryptographic protocol's implementation is secure in the non-ideal environment.

PROOF: Since there are not logic flaws in $\beta_1, \beta_2, \cdots \beta_i \cdots \beta_j \cdots \beta_n$ , $\beta_1, \beta_2, \cdots \beta_i \cdots \beta_j \cdots \beta_n$ can satisfy judgement I and judgement II of the DFLSAF., and the KMP( $\beta_1, \beta_2, \cdots \beta_i \cdots \beta_j \cdots \beta_n$ , $\beta'_1, \beta'_2, \cdots \beta'_i \cdots \beta'_j \cdots \beta'_n$ , pos) is matched successfully, so that the cryptographic protocol's implementation is secure in the non-ideal environment.

According to semantically equivalent definition, the semantic of $\beta_1, \beta_2, \cdots \beta_i \cdots \beta_j \cdots \beta_n$ is equal to .the semantic of $\beta'_1, \beta'_2, \cdots \beta'_i \cdots \beta'_j \cdots \beta'_n$ . Since $\beta'_1, \beta'_2, \cdots \beta'_i \cdots \beta'_j \cdots \beta'_n$ satisfies judgement I and judgement II of DFLSAF, $\beta_1, \beta_2, \cdots \beta_i \cdots \beta_j \cdots \beta_n$ also satisfies judgement I and judgement II of DFLSAF. Therefore, the cryptographic protocol's implementation is secure in the non-ideal environment.

5. The experiment

In order to verify how to apply $L\pi CT$ to the logic security analysis of cryptographic protocols, in this paper TLS1.3 standard, officially published in 2018, is taken to empirically analyze the logic security of TLS1.3 protocol interactional process.

5.1 TLS1.3 and its Interactional process

Based on TLS1.2[32], TLS1.3 was improved by IETF (Internet Engineering Task Force) four times, and was published in 2018.

1. The structure of TLS1.3 protocol

Derived from SSL, TLS1.3 has been improved on the base of TLS1.2, and has added round-trip time (0-RTT) so as to reduce the time delay of the Interactional communication. Its structure includes recording protocol, handshake protocol and warning protocol. In TLS1.3, the handshake protocol has deleted the regulation of the updating password specification. The Interactional process of TLS1.3 is shown in 5-1.

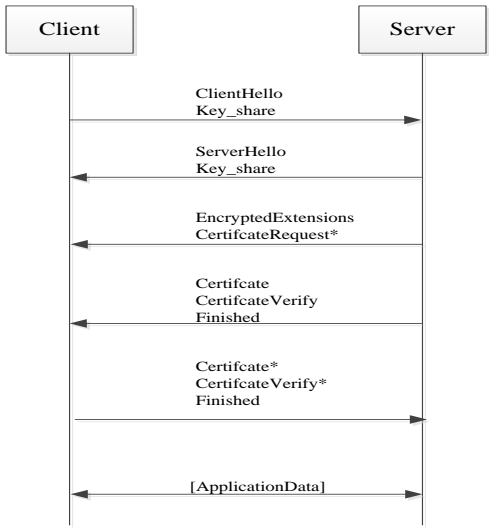

Figure 5-1. A complete TLS1.3 handshake protocol

The handshake protocol of TLS1.3 needs to extend sharing key, sending authentication messages and so on. Figure 5-1 only describes the basic Interactional process of TLS1.3 protocol.

  2.The interactional process of TLS1.3

  Compared to TLS1.2, TLS1.3 has simplified the Interactional process, reducing the time delay of Interactional communication and strengthening the confidentiality when transiting information. First, in the procedure of Interactional handshake TLS1.3 supports three kinds of key exchange: ① the key exchange of Diffie-Hellman protocol based on a finite field group or elliptic curve; ②PSK（pre-share key), the key exchange employing pre-sharing； ③ the pre-share key of Diffie-Hellman protocol. Second，the handshake protocol of TLS1.3 includes three stages: the first stage is to key exchange, and on this stage two communication parties create the information by sharing the keys and choose their own key parameters ;on the second stage two communication parties create handshake parameters; The third stage is authentication and on this stage two communication parties finish mutual authentication and the integrity verification of sending or receiving messages. At last, two communication parties share data application. Specific process details are shown in figure 5-2.

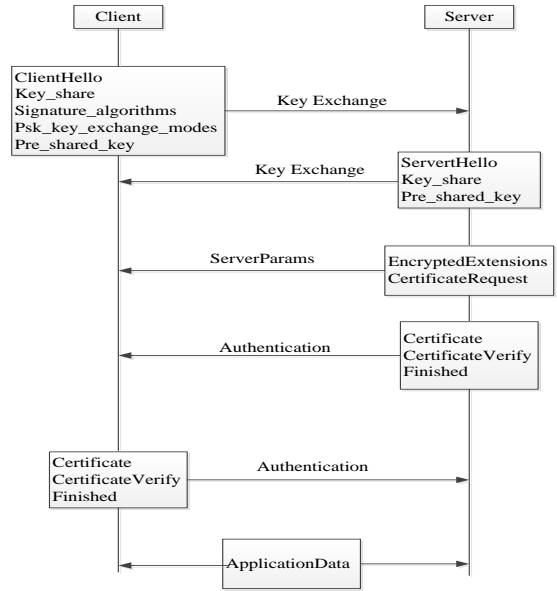

Figure 5-2. the Interactional process details of TLS1.3 protocol

  3.Interactional steps

  Step1. clients send ClientHello to the server，which includes a random nonce, protocol version, a symmetric key or a pair list HKDF , a group of Diffie-Hellman key_share and a group of pre_shared_key labels（pre_shared_key）, and so on.

  Step 2. the server receives the message ClientHello sent by clients, and chooses the encrypted parameters connected to the clients. After that, the server sends ServerHello to clients. "Key_share" includes the temporary Diffie-Hellman key_share of the server. The server's key_share must be in the same group with the clients' key_share. If the server and the clients are using PSK keys to create the connection between them, ServerHello includes a "pre_shared_key"extension. The"pre_shared_key"extension indicates which clients provide chosen PSK.

  Step 3. the server sends two messages to the clients: Encrypted Extensions and Certificate Request，which are involved in certificate and the authentication of the clients.

  Step 4. The server sends Certificate and Certificate Verify to the clients. If the server does not use the certificate to authenticate, the server will omit this message. If the clients don't use the certificate to authenticate, this message will be omitted by the clients.

  Step 5. Upon receiving the message from the server, the clients respond with its authentication.

  The clients and the servers execute step1 to step5, and it indicates that the handshake protocol has been completed, and that the clients and the server have obtained the key information which is needed when the recording layer encrypt the data of the applying layer.

 5.2TLS1.3 Logic security analysis of $L\pi CT$

  Logic rules and proofs are added to $L\pi CT$ , which is based on π- calculation theory, in order to analyze whether there are logic flaws in the states transition of a protocol's Interactional communication, so that attacks will be reduced during the protocol implementation. First, the states transition is symbolized in the Interactional communication of TSL1.3 . Second, logic rules and proofs are used to analyze whether there are logic flaws in TSL1.3. In this paper, the attack capability of an adversary is introduced as a precondition to analyze the logic security of TLS1.3 as follows:

5.3 Symbolization in the interaction process of TLS1.3

1. A round of handshake interaction process of TLS1.3

During the implementation of TLS1.3, the nodes are labelled in a round of the handshake interactions as follows in figure 5-3.

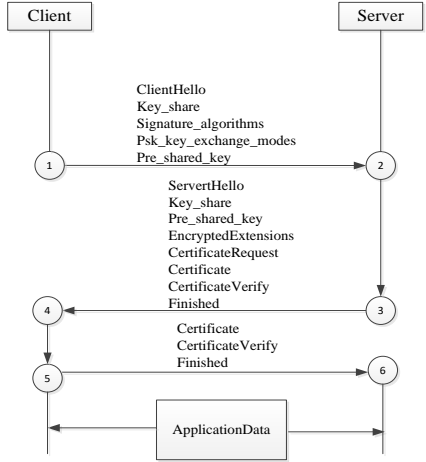

Figure 5-3 Node labelling

2. Building the symbol model in a round of the handshake interaction of TLS1.3

As shown in figure 5-2, there are 6 nodes in a around of the handshake interaction process of TLS1.3, marked as ①, ②, ③、 ④, ⑤, ⑥. The security of the corresponding events guarantees the security of a protocol's handshake. Therefore, each node has to finish corresponding events before carrying out the handshake process with other nodes.

3. The states transition system of TLS1.3 based on π-calculation

According to Definition 1, the states transition system of TLS1.3 is a two-tuples $(Q, T)$. Here, $Q$ denotes a states set: $Q \subseteq \{S_i \mid 1 \leq i \leq 6, i \in N\}$; $T$ denotes a triple: $T \subseteq (Q \times msg \times Q)$; $msg$ denotes the messages of a node's corresponding events: $msg = \{msg_j, 1 \leq j \leq 6\}$. For example, the message $msg_1$ of the corresponding event of the states $S_1$ includes {ClientHello, Key_share, Signature_algorithms, Psk_key_exchange_modes, Pre_shared_key}. The messages of other states can also be represented in this way. According to definition 1 and figure 5-3, the states transition figure of TSL1.3 is built as follows in figure 5-4.

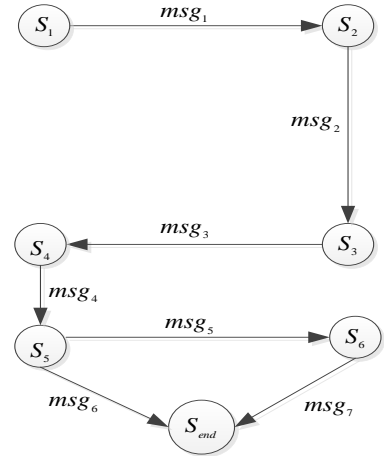

Figure 5-4. the states transition of TLS1.3

Here, $msg_1 = \begin{Bmatrix} ClientHello \\ Key\_share \\ Signature\_algorithms \\ Psk\_key\_exchange\_modes \\ pre\_shared\_key \end{Bmatrix}$,

$msg_2 = msg_3 = \begin{Bmatrix} ServertHello \\ Key\_share \\ Pre\_shared\_key \\ EncryptedExtensions \\ Certificate\,Request \\ Certificate \\ CertificateVerify \\ Finished \end{Bmatrix}$

$msg_4 = msg_5 = \begin{Bmatrix} Certificate \\ CertificateVerify \\ Finished \end{Bmatrix}$ and

$msg_6 = msg_7 = \{ApplicationData\}$

(1) The logic proposition entailment and proof of the states transition of TLS1.3

As is shown in figure5-3 and figure 5-4, the states transition of TLS1.3 is an entailment relation between premise and conclusion. There is also an entailment relation between every states corresponding event and states transition. The states transition of TLS1.3 can be represented as a two-tuples $(S, \rightarrow)$. Here, $S$ is a states transition set; $\rightarrow$ is states transition relation and $(S, \rightarrow)$ satisfies partial ordering relation.

Logic entailment can be used to prove whether is true or not, and judge whether there are logic flaws in the TLS1.3. Here, the antecedent of sequent(precondition) is $S_1 \rightarrow S_2, S_2 \rightarrow S_3, S_3 \rightarrow S_4, S_4 \rightarrow S_5, S_5 \rightarrow S_6$ ; and the consequent(conclusion) is : $S_{end}$ .

①Logic entailment: prove that sequent $S_1 \rightarrow S_2, S_2 \rightarrow S_3, S_3 \rightarrow S_4, S_4 \rightarrow S_5, S_5 \rightarrow S_6, S_6 \rightarrow S_{end} \mid - S_{end}$ is valid.

**PROOF:** Consider two cases in the logic security analysis of TLS1.3, the ideal environment and the non-ideal environment.

②the ideal environment

According to the entailment elimination rules of logic rules:

$$\frac{\phi \quad \phi \rightarrow \psi}{\psi} \rightarrow i \qquad (5\text{-}1)$$

---

**logic entailment and proof can be written as follows:**

1. $S_1$      promise
2. $S_1 \rightarrow S_2$      assumption
3. $S_2$      $\rightarrow e$   2,1
4. $S_2 \rightarrow S_3$      assumption
5. $S_3$      $\rightarrow e$   4,3
6. $S_3 \rightarrow S_4$      assumption
7. $S_4$      $\rightarrow e$   6,5
8. $S_4 \rightarrow S_5$      assumption
9. $S_5$      $\rightarrow e$   8,7
10. $S_5 \rightarrow S_6$      assumption
11. $S_6$      $\rightarrow e$   10,9
12. $S_6 \rightarrow S_{end}$      assumption
13. $S_{end}$      $\rightarrow e$   12,11(conclusion)

---

Explanation: first, $S_1$ is the original states of TLS1.3 and $S_{end}$ is the conclusion of TLS1.3. Second, the "assumption" of the proof is certain to happen in the ideal environment. Finally, in the ideal environment, with assumption( states transition), it can be deduced that $S_{end}$ is true.

According to logic reduction proof, there are not logic flaws in the ideal environment the interaction process of TSL1.3.

Over □

③ the non-ideal environment

In the non-ideal environment, it is assumed that attackers have the capabilities of Dolev-Yao model. In this case, logic entailment is used to prove the logic security in the interaction process of TLS1.3 .

In the non-ideal environment, it is difficult to use positive order proof to prove whether there are logic flaws in the interaction process of TLS1.3. In this paper, the proof by contradiction is applied. Starting from the conclusion $S_{end}$ of TLS1.3, the assumption is proposed that $\neg S_{end}$ is true, which is deducted to be false. In this way, it is known that $\neg S_{end}$ is not true. Therefore, the opposite of this assumption is true. That is, the conclusion $S_{end}$ is true.

According to the rules of the proof by contradiction

$$\frac{\phi \rightarrow \psi, \quad \neg \psi}{\neg \phi} \quad MT \qquad (5\text{-}2)$$

---

**logic entailment and proof can be written as follows:**

1. $\neg S_{end}$      assumption
2. $S_6 \rightarrow S_{end}$      premise
3. $\neg S_6$      MT 2,1
4. $S_5 \rightarrow S_6$      premise
5. $\neg S_5$      MT 4,3
6. $S_4 \rightarrow S_5$      premise
7. $\neg S_4$      MT 6,5
8. $S_3 \rightarrow S_4$      premise
9. $\neg S_3$      MT 8,7
10. $S_2 \rightarrow S_3$      premise
11. $\neg S_2$      MT 10,9
12. $S_1 \rightarrow S_2$      premise
13. $\neg S_1$      (over) MT 11,12
14. $S_1$      premise
15. $\perp$      $\neg e$ ( $\perp$ denotes contradiction)

---

As is proved by logic rules above, the assumption is false. As a result, there are not logic flaws in the TLS1.3 interaction implementation.

over□

According to the rules of the proof by contradiction, attackers have the capabilities of Dolev-Yao model . In the interaction process, every transition states of TLS1.3 is vulnerable to attacks, which is in line with what happens when TLS1.3 is implemented in the open cyberspace. The design and the implementation of TSL1.3 are secure only when each node's states $S_i ( 2 \leq i , i \in N )$ is secure and the logic value of each corresponding event is true.

(2)Logic security analysis of the corresponding events of each states node

According to figure 5-3 and figure 5-4, the corresponding events of each states node can be written as follows in table 1:

Table 1 the corresponding events of each states node

| Node name | Corresponding events |
|---|---|
| $S_1$ | 1.*ClientHello*, 2. *Key_share*, 3.*Signature_algorithms*, 4.*Psk_key_exchange_modes*, 5.*Pre_shared_key* |
| $S_2 = S_3$ | 1.*ServerHello*, 2.*Key_share*, 3.*Pre_shared_key*, 4.*EncryptedExtensions*, 5.*Certificate* Re *quest*, 6.*Certificate*, 7.*CertificateVerify*, 8.*Finished* |
| $S_4 = S_5$ | 1.*Certificate*, 2.*CertificateVerify* |
| $S_6 = S_7$ | *ApplicationData* |

Assume that the states transition node $S_i$ has $n$ corresponding events. The security of $S_i$ depends on the logic conjunctive computation of each states transition node's corresponding events, which is written as follows:

$$S_i = e_1 \wedge e_2 \wedge \cdots \wedge e_i \qquad (5\text{-}3)$$

According to the equation 5-3, only when the value of every corresponding event is true, will the value of each states transition node be true. Under such condition, the security of each states transition node can be ensured.

(3)the security analysis of corresponding events

To analyze the logic security of corresponding events, this paper takes $S_1$ for example and builds a binary tree for its corresponding events, shown in Figure 5-5.

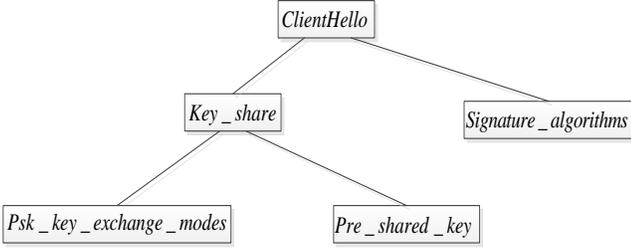

Figure 5-5 The binary tree for the corresponding events of states transition nodes $S_1$

According to DFLSAF algorithm, an conjunctive computation equation can be written for the corresponding events of the states transition node $S_1$ in TLS1.3:

$$S_1 = ClientHello \land Key\_share \\ \land Signature\_algorithms \land \\ Psk\_key\_exchange\_modes \land \\ Pre\_shared\_key \quad (5\text{-}4)$$

Therefore, $S_1$ is secure only when the conjunctive computation value of the corresponding events of $S_1$ in TLS1.3 is true. That is, there are not logic flaws in $S_1$. Similarly, other states transition nodes $S_2, S_3, S_4, S_5, S_6, S_7$ can build binary trees of their corresponding events and carry out conjunctive computation value. There are not logic flaws in each states transition node only when the conjunctive computation value of their corresponding events is true. Therefore, the security of TLS1.3's design and implementation is related to the corresponding events, shown in Table 2.

Table 2 Logic relationship between states transition nodes and the corresponding events

| Node name | Corresponding events |
|---|---|
| $S_1$ | $ClientHello \land Key\_share \land$ $Signature\_algorithms \land$ $Psk\_key\_exchange\_modes \land$ $Pre\_shared\_key$ |
| $S_2 = S_3$ | $ServertHello \land Key\_share \land$ $Pre\_shared\_key \land$ $EncryptedExtensions \land$ $Certificate\,\mathrm{Re}\,quest \land$ $Certificate \land$ $CertificateVerify \land Finished$ |
| $S_4 = S_5$ | $Certificate \land CertificateVerify$ |
| $S_6 = S_7$ | $ApplicationData \lor \neg ApplicationData$ |

As is shown in Table 2, the corresponding events $Signature\_algorithms$ of $S_1$ are involved in signature algorithms. Signature algorithms are involved in symmetric or asymmetric encryption and decryption, which are involved in public and private keys. And public and private keys are involved in difficult mathematical problems (large integer factorization, elliptic curves, and discrete logarithms). That is to say, their logic security decides the security of the corresponding events $Signature\_algorithms$

The security of TLS1.3's implementation is related to the security of each states transition node, and the security of each states transition node is decided by its corresponding events. In order to clearly understand the security of $S_i$ ( $2 \leq i$, $i \in N$ ) in TLS1.3 and the security of the corresponding events, the security association with table between them is shown in Table 3.

Table 3 the association between the security of states transition nodes and the security of corresponding events

| Node name | Corresponding events | the type resistant to attack or providing security |
|---|---|---|
| $S_1$ | $ClientHello$, $Key\_share$, $Signature\_algorithms$, $Psk\_key\_exchange\_modes$, $Pre\_shared\_key$ | resisting replay attacks, resisting man-in-the-middle attacks, ensuring forward safety, integrity, identity authentication, ensuring Client and Server selection synchronization |
| $S_2$ | $ServertHello$, $Key\_share$, $Pre\_shared\_key$, $EncryptedExtensions$, $CertificateRequest$, $Certificate$, $CertificateVerify$, $Finished$ | resisting replay attacks, resisting man-in-the-middle attacks, ensuring forward security, integrity, identity authentication, ensuring Client and Server selection synchronization, protecting information confidentiality, verification. |
| $S_4$ | $Certificate$, $CertificateVerify$ | providing the information verification function to ensure its authenticity |
| $S_6$ | $ApplicationData$ | verifying the protocol security implementation |

In table 3, $S_2 = S_3$; $S_4 = S_5$ and $S_6 = S_7$.

As is shown in Table 3, each states transition node has its corresponding events, the security of which decide the security of TLS1.3's design and implementation. Therefore, when a protocol is being designed or implemented, the logic security of every states transition node's corresponding events should be analyzed in order to avoid logic flaws in the design and implementation, so as to protect the protocol from attacks when it is encoded into a software.

In the ideal environment, there is no logic defect in the sequent of TLS1.3 protocol interactional communication, and the perfect match between the sequent of TLS1.3 protocol interactional communication in the ideal environment and that of TLS1.3 protocol interactional communication in the non-ideal environment is obtained by calling KMP algorithm. According to theorem 1, there is no logic defect in TLS1.3 protocol interactional communication in the non-ideal environment. According to the calculation results, the sequent $\{\{(S_i, \rightarrow) | 1 \le i \le 7, i \in N\}$ of TLS1.3 protocol states transition node satisfy the judgment III of DFLSAF algorithm and theorem 1. The conclusion is that there is no logic defect in the design and interaction of TLS1.3 protocol.

In the ideal environment, the logic entailment of TLS1.3's interactional communication sequent are effective, so there is no logic flaws in the design and the implementation of TLS1.3 in the ideal environment. Through KMP algorithm, it is found that the two sequent, obtained by TLS1.3 in two different environments (the ideal environment and the non-ideal environment), are exactly matched. According to Theorem 1, there are not logic flaws in the Interactional communication of TLS1.3 in the non-ideal environment. Therefore, TLS1.3's interactional communication sequent satisfy the judgments I, II and III of the DFLSAF. That is, there are not logic flaws in the design and the implementation of TLS1.3.

Similarly, DFLSAF algorithm, based on $L\pi CT$, is also suitable for the logic security analysis of other protocols, such as Diffie-Hellman protocol, Schnorr protocol, MQV protocol, Kerbers protocol and so on. DFLSAF algorithm can be used to judge that Diffie-Hellman protocol can not resist man-in-the-middle attack, impersonation attack and so on. When Diffie-Hellman protocol is encoded into a software and implemented, random number will be attacked. Therefore, Diffie-Hellman protocol is not resistant to replay attacks, and can not guarantee forward security. Due to the limited space, other protocols will not be discussed here.

## 6. Discussion

In this paper, TLS1.3 protocol is taken as an example, and its interaction process is marked and symbolized, and the logic calculation relationship table between the states transition nodes and the corresponding events is constructed. Logic rules are used to prove the logic security of TLS1.3 in two different environments. The deficiencies of Diffie-Hellman protocol are also analyzed. The empirical results show that the new theory, framework and algorithm are feasible and effective for analyzing the logic security of a cryptographic protocol's design and implementation, and are suitable for the logic security entailment and proof of a cryptographic protocol's design and implementation.

The future trend of this work is to apply mixed-cross methods such as machine learning, knowledge map and statistic analysis, etc. In different network and applying environments and on different operating system platforms to carry out dynamic modeling and logic provable security analysis when encoding a cryptographic protocol into a software and implementing it.

## 7. Conclusion

The logic security of a cryptographic protocol's design and implementation is the security basis when it is encoded into a software and implemented. In the field of cyberspace security, it is a difficult problem, as well as one of the important researches, to analyze whether there are logic flaws in the design and implementation of a cryptographic protocol. In order to solve this problem, this paper, by introducing logic rules and logic proof, extends the π-calculus theory, and proposes the theory LπCT and DFLSAF algorithm.

This paper introduces the definition of sequent in the logic security entailment and proof and proves the validity of sequent in the cryptographic protocol's design and implementation. In this paper, the binary trees of states transition nodes and the binary trees of corresponding events are built, which makes it convenient and easy to encode the process of designing and implementing a cryptographic protocol, and search states transition nodes. In order to ensure DFLSAF to be correct, KMP algorithm is used to match sequent obtained in two different environments during designing and implementing the cryptographic protocol.

## Declaration of Competing Interest

The authors declare that they have no known competing financial interests or personal relationships that could have appeared to influence the work reported in this paper.

## Acknowledgment

The authors are Supported by National Science Foundation of China (No.62062019, 62061007, 62072247) , Guizhou University of Finance and Economics Innovation Exploration and Academic New Seedling Project (No.2022XSXMA11) and 2020 Provincial Department of Education Natural Science Research Project (No. Qianjiaohe KYzhi[2021]064) .

## References

Milner R . Communicating and Mobile Systems : The π-calculus[M]. 1999.
Michael HuthMark ， RyanMark Ryan. Logic in Computer Science: Modelling and Reasoning About Systems(second edition).Cabridge University Press,2004
Monin J F , Mgh C . Classical Logic[M]. Springer London, 2003
MICHAEL HUTH ， MARK RYAN. LOGIC IN COMPUTER SCIENCE-Modelling and Reasoning about Systems. Cambridge University Press 2004
David Baelde, Stéphanie Delauney, Charlie Jacomme, et al. An Interactional Prover for Protocol Verification in the Computational Model. 42 nd IEEE Symposium on Security and Privacy. may 24-27, 2021, take place virtually.
Petersen L , Pontelli E .Declarative aspects of multicore programming: Foreword[J]. 2010.
Ross Horne. A Bisimilarity Congruence for the Applied pi-Calculus Sufficiently Coarse to Verify Privacy Properties. arXiv:1811.02536v1.2018
Cheval V , Kremer S , LORIA, et al. DEEPSEC: Deciding Equivalence Properties in Security Protocols Theory and Practice[C]// 2018 IEEE Symposium on Security and Privacy (SP). IEEE, 2018.
Bodei C , Brodo L , Bruni R . The link-calculus for Open Multiparty Interactions[J]. Information and Computation, 2020, 275:104587.
Bruno Blanchet.Modeling and Verifying Security Protocols with the Applied Pi Calculus and ProVerif. Foundations and Trends,2016,2:1-135
Thomas Troels Hildebrandt,Christian Johansen, Håkon Normann. A stable non-interleaving early operational semantics for the pi-calculus. Journal of Logic and Algebraic Methods in Programming 104 (2019) 227–253
Kokke W , Montesi F , Peressotti M . Better Late Than Never: A Fully Abstract Semantics for Classical Processes[J]. 2018. arXiv:1811.02209

Niccolò Veltri. Formalizing the π-calculus in Guarded Cubical Agda. In Proceedings of the 9th ACM SIGPLAN International Conference on Certifed Programs and Proofs (CPP '20), January 20–21, 2020, New Orleans, LA, USA. ACM, New York, NY, USA, 14 pages

Martín Abadi，Bruno Blanchet，Cédric Fournet，Authors Info & Affiliations. The Applied Pi Calculus: Mobile Values, New Names, and Secure Communication. Journal of the ACM, 2017,Vol. 65, No. 1

Livinus Obiora Nweke, Goitom K. Weldehawaryat,Stephen D. Wolthusen. Threat Modelling of Cyber–Physical Systems Using an Applied the π-calculus. international journal of critical infrastructure protection 35 (2021) 100466

Yi Chen，Luyi Xing, Yue Qin，et al. Devils in the Guidance: Predicting Logic Vulnerabilities in Payment Syndication Services through Automated Documentation Analysis. 28th USENIX Security Symposium. August 14– 16, 2019 ,Santa Clara, CA, USA

Edward J. Schwartz，Cory F. Cohen，Michael Duggan，et al. Using Logic Programming to Recover C++ Classes and Methods from Compiled Executables. CCS'18, October 15-19, 2018, Toronto, ON, Canada

Nigam, V., Ban Kirigin, T., Talcott, C，et al. Logic, Language, and Security. Springer Nature Switzerland AG 2020

Jean Goubault-Larrecq,Fabrice Parrennes. Cryptographic Protocol Analysis on Real C Code.VMCA2005:proceedings of the International Conference on Verifcation, Model Checking and Abstract Interpretation.2005[C].Paris, France, July 6-8.

Jia Chen，Yu Feng，Isil Dillig. Precise Detection of Side-Channel Vulnerabilities using Quantitative Cartesian Hoare Logic[C]// the 2017 ACM SIGSAC Conference. ACM, 2017

Anindya Banerjee, David A. Naumann, and Mohammad Nikouei. A Logic Analysis of Framing for Specifications with Pure Method Calls. ACM Trans. Program. Lang.2018

Banerjee A , Nagasamudram R , Naumann D A , et al. A Relational Program Logic with Data Abstraction and Dynamic Framing. arXiv:1910.14560,2019

Ribeiro M , Ado P , Mateus P . Formal Verification of Ethereum Smart Contracts Using Isabelle/HOL[M]. Logic, Language, and Security，

Tao Liu, Yangjia Li, Shuling Wang, Mingsheng Ying, and Naijun Zhan.2016. A theorem prover for quantum Hoare logic and its applications. arXiv preprint arXiv:1601.03835 (2016). https://arxiv.org/abs/1601.03835

Li Zhou, Nengkun Yu, and Mingsheng Ying. 2019. An Applied Quantum Hoare Logic. In Proceedings of the 40th ACM SIGPLAN Conference on Programming Language Design and Implementation (PLDI '19), June 22–26, 2019